\documentclass[conference]{IEEEtran}
\IEEEoverridecommandlockouts
\usepackage{cite}
\usepackage{amsmath,amssymb,amsfonts}
\usepackage{algorithmic}
\usepackage{graphicx}
\usepackage{textcomp}
\usepackage{xcolor}
\usepackage{tabularx}

\def\BibTeX{{\rm B\kern-.05em{\sc i\kern-.025em b}\kern-.08em
    T\kern-.1667em\lower.7ex\hbox{E}\kern-.125emX}}
\begin{document}

\title{Advancing Cyber-Attack Detection in Power Systems: A Comparative Study of Machine Learning and Graph Neural Network Approaches
\thanks{This work was supported by the U.S. Department of Energy’s (DOE) Office of Cybersecurity, Energy Security, and Emergency Response (CESER) and performed at the Pacific Northwest National Laboratory (PNNL), operated for the U.S. DOE by Battelle Memorial Institute under Contract No. DE-AC05-76RL01830.}
}

\author{\IEEEauthorblockN{Tianzhixi Yin}
\IEEEauthorblockA{\textit{AI \& Data Analytics} \\
\textit{PNNL}\\
Richland, WA, USA \\
tianzhixi.yin@pnnl.gov}
\and
\IEEEauthorblockN{Syed Ahsan Raza Naqvi}
\IEEEauthorblockA{\textit{Electricity Infrastructure} \\
\textit{PNNL}\\
Richland, WA, USA \\
ahsan.raza@pnnl.gov}
\and
\IEEEauthorblockN{Sai Pushpak Nandanoori}
\IEEEauthorblockA{\textit{Electricity Infrastructure} \\
\textit{PNNL}\\
Richland, WA, USA \\
saipushpak.n@pnnl.gov}
\and
\IEEEauthorblockN{Soumya Kundu}
\IEEEauthorblockA{\textit{Electricity Infrastructure} \\
\textit{PNNL}\\
Richland, WA, USA \\
soumya.kundu@pnnl.gov}

}

\maketitle

\begin{abstract}
This paper explores the detection and localization of cyber-attacks on time-series measurements data in power systems, focusing on comparing conventional machine learning (ML) like k-means, deep learning method like autoencoder, and graph neural network (GNN)-based techniques. We assess the detection accuracy of these approaches and their potential to pinpoint the locations of specific sensor measurements under attack. Given the demonstrated success of GNNs in other time-series anomaly detection applications, we aim to evaluate their performance within the context of power systems cyber-attacks on sensor measurements. Utilizing the IEEE 68-bus system, we simulated four types of false data attacks, including scaling attacks, additive attacks, and their combinations, to test the selected approaches. Our results indicate that GNN-based methods outperform k-means and autoencoder in detection. Additionally, GNNs show promise in accurately localizing attacks for simple scenarios, although they still face challenges in more complex cases, especially ones that involve combinations of scaling and additive attacks.
\end{abstract}


\section{Introduction}


Cyber-attacks on power systems can have devastating effects, disrupting essential services and causing significant economic losses \cite{Xu}. As power systems become increasingly interconnected and digitized, they become more vulnerable to sophisticated cyber-attacks, as demonstrated by real cyber-attacks \cite{case2016analysis,reilly2015bracing}. Detecting cyber-attacks on power systems in a timely manner is of paramount importance because it allows for swift mitigation measures, minimizing the impact of the attack and ensuring the resilience and reliability of the power grid \cite{Liang,He}. Increasing cloud-based communication and control systems, fast-evolving ransomware threats, and the convergence of information technology (IT) and operations technology (OT) systems were identified as among the emerging cybersecurity challenges faced by the power grid in a recent report by the US Department of Energy \cite{us2022cybersecurity}. Most existing grid cybersecurity solutions either focus only on IT-based intrusion detection \cite{chan2022,singh2022}, or a hybrid (combined IT/OT) intrusion detection but with static, pre-determined, OT rules \cite{johnson2022,eos}. While rule-based cyber intrusion detection engines rely on operators' OT knowledge to establish a `baseline', these suffer from a lack of adaptability, especially as new energy resources with evolving controls and communications are integrated into the grid at an increasing rate \cite{almassalkhi2023intelligent}.


There has been various research focusing on OT-based automated cyber-attack detection for power systems using different algorithmic (non-rule-based) methods, including model-based, machine learning, and deep learning methods. A broad class of the cyber-detection methods use physics-based models with an outlier detection algorithm \cite{ghosal2018diagnosis,murguia2016characterization,huang2018online}. Different machine learning-based methods, on the other hand, have deployed time-series modeling approach \cite{Abedi}, decision trees \cite{Vedant}, ensemble methods \cite{Lu}, or unsupervised k-nearest neighbor approach \cite{Bouyeddou}. More recently, various deep learning-based intrusion detection approaches are being investigated, e.g., the Convolutional Neural Network (CNN) approach \cite{Li}, deep reinforcement learning \cite{SPADES}.

In this research, we study cyber-attacks that are injected into the sensor measurements of voltage angles \cite{Ahmed}. We look specifically at the cases when attacks are introduced right after a grid event, where the attacks can be hidden behind transient disturbances. This study explores the application of several machine learning (ML) techniques to detect cyber-attacks on power systems, not only identifying the occurrence of an attack but also investigating the specific buses that are under attack. Traditional ML methods like k-means clustering \cite{MacQueen} and deep learning-based methods such as autoencoders \cite{Hinton} combined with k-means were tested in this study. However, the complex and interconnected nature of power systems suggests that graph neural network (GNN) approaches, including Graph Attention Networks (GAT) \cite{Zhao} and Graph Deviation Networks (GDN) \cite{Deng}, might offer superior detection capabilities. This paper details our investigation into these methods, comparing their effectiveness in the simulated environment of the IEEE 68-bus system.


\section{Methodology}

Our study simulated four types of cyber-attack scenarios on a power system model to generate a diverse dataset for analysis. These scenarios included Step, Data poisoning, Ramp, and Riding the Wave (RTW). We employed a variety of ML methods to analyze the dataset, aiming to detect the occurrence of cyber-attacks and identify the specific buses under attack. The cyber-attack detection problem is usually a multivariate time series anomaly detection problem. We began with conventional ML methods, utilizing k-means clustering to segment the time series for data indicative of normal operation and those suggestive of cyber-attacks. Next, we explored a deep learning-based approach, implementing an autoencoder to learn the normal operational patterns of the power system. We then applied the same k-means clustering to the latent space representations generated by the autoencoder, aiming to distinguish between normal and attack scenarios.

Pazho et al. (2023) \cite{Pazho} conducted a comprehensive survey of graph-based deep learning methodologies for anomaly detection in distributed systems. Based on the availability of code, data similarity to power system time-series data, and with an emphasis on transformer-like attention mechanisms, we selected GAT and GDN for our dataset evaluation. These two methods leverage the inherent graph structure of power systems, where nodes represent buses and edges represent transmission lines. GAT focus on using attention layers to capture both the dependencies in the temporal and feature spaces among different time series. It combines forecasting and reconstruction losses to determine the abnormality of multivariate time series. Meanwhile, GDN aim to learn the dependence relationships between buses using attention-based graph neural network, and detect deviations from these relationships. Unlike GAT, GDN uses only the forecasting error.


\subsection{Attack Simulation}
\par
\begin{figure}[htbp]
\centerline{\includegraphics[width=0.9\columnwidth]{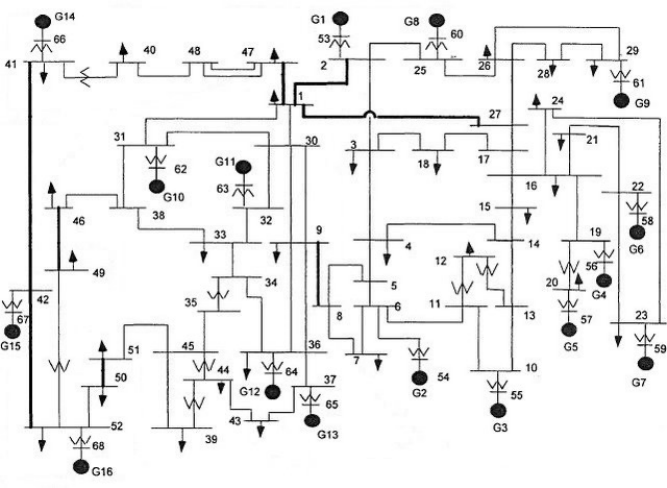}}
\caption{Single line diagram of the IEEE 68-bus system with 16 SGs.}
\label{fig:IEEE68Bus}
\end{figure}

 In this work, we consider a MATLAB-based framework that simulates the power system dynamics. We consider the IEEE $68$-bus system (see Fig. \ref{fig:IEEE68Bus}) consisting of $16$ synchronous generators (SGs) grouped into 5 coherent areas. The SGs are modeled with 1-axis governor controls which acts as primary controls for frequency regulation. The system is assumed to be equipped with sensors such as PMUs at each bus, measuring voltage angle, voltage magnitude, frequency and change in frequency. The secondary control, automatic generation control (AGC) is also implemented using the wide-area measurements. The simulation framework has the capability to model grid events such as load changes, generator loss, line tripping and faults. 
\par In this paper, we consider an adversarial scenario where an attacker injects attack signals into the sensor measurements for voltage angles. A few comments regarding this assumption are in order. Typically, an adversary would have access to all measurements in a PMU. However, since this work considers real power flow only, injecting attack signals into the voltage magnitude measurements would have minimal impact on the system behavior. Frequency measurements, in contrast, are essential for grid operations as they, in turn, can determine the generation levels of SGs in subsequent time instances. Hence, injecting attack signals into frequency measurements can have a significant, adverse impact on system performance. However, attacking voltage angle measurements can be detrimental in two aspects. Firstly, it can impact the estimates of the real power flow between two buses. Secondly, since the change in voltage angles is used to determine the instantaneous frequency, an attack on angle measurements can introduce significant aberrations in the behavior of the AGC. Therefore, given the importance of voltage angles, this work assumes that the attacker introduces attack signals into these measurements. The attacked measurements when passed to AGC for frequency restoration, will result in undesirable set-point changes by the AGC and the objective of frequency restoration is not achieved. Therefore, it is imperative to detect these attacks that hamper the operation of the power system which forms the scope of this work. 
\par In every attack strategy described below, when the simulation begins, the system is in steady state and at time $1$s, there is a load change. Then, at time $2$s, when the system is reacting to the load change, there is an attack and the attack lasts until $22$s. Specifically, we consider the cases where an attack is introduced just after a grid event (such as a load change or generation change), so that the attack signal is hidden behind existing transient disturbances in the power grid. 
\par Similar to the types of attacks presented in \cite{Huang2009}, the attack strategies studied in this work can be categorized into (i) scaling attacks (step, ramp), (ii) additive attacks (poison), and (iii) a combination of both (RTW). We will now summarize the different types of attacks studied in the scope of this work. 
\subsubsection{Step attack}
\par
\begin{figure}[htbp]
\centerline{\includegraphics[width=0.95\columnwidth]{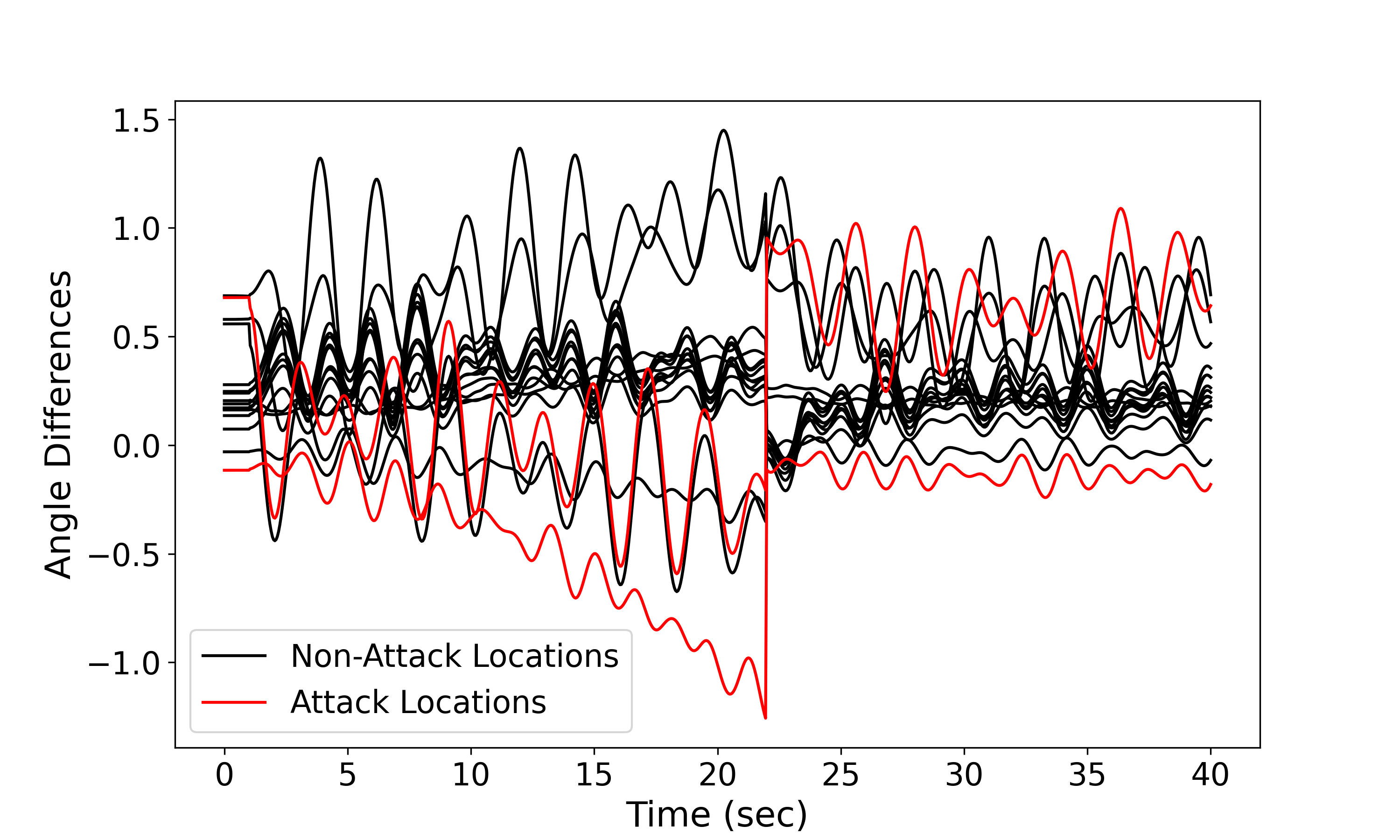}}
\caption{Step scenario with small attack magnitude with $c=1.006$. The attack results in gradual divergence of angle differences during the attack.}
\label{sce46_step}
\end{figure}

\begin{figure}[htbp]
\centerline{\includegraphics[width=0.95\columnwidth]{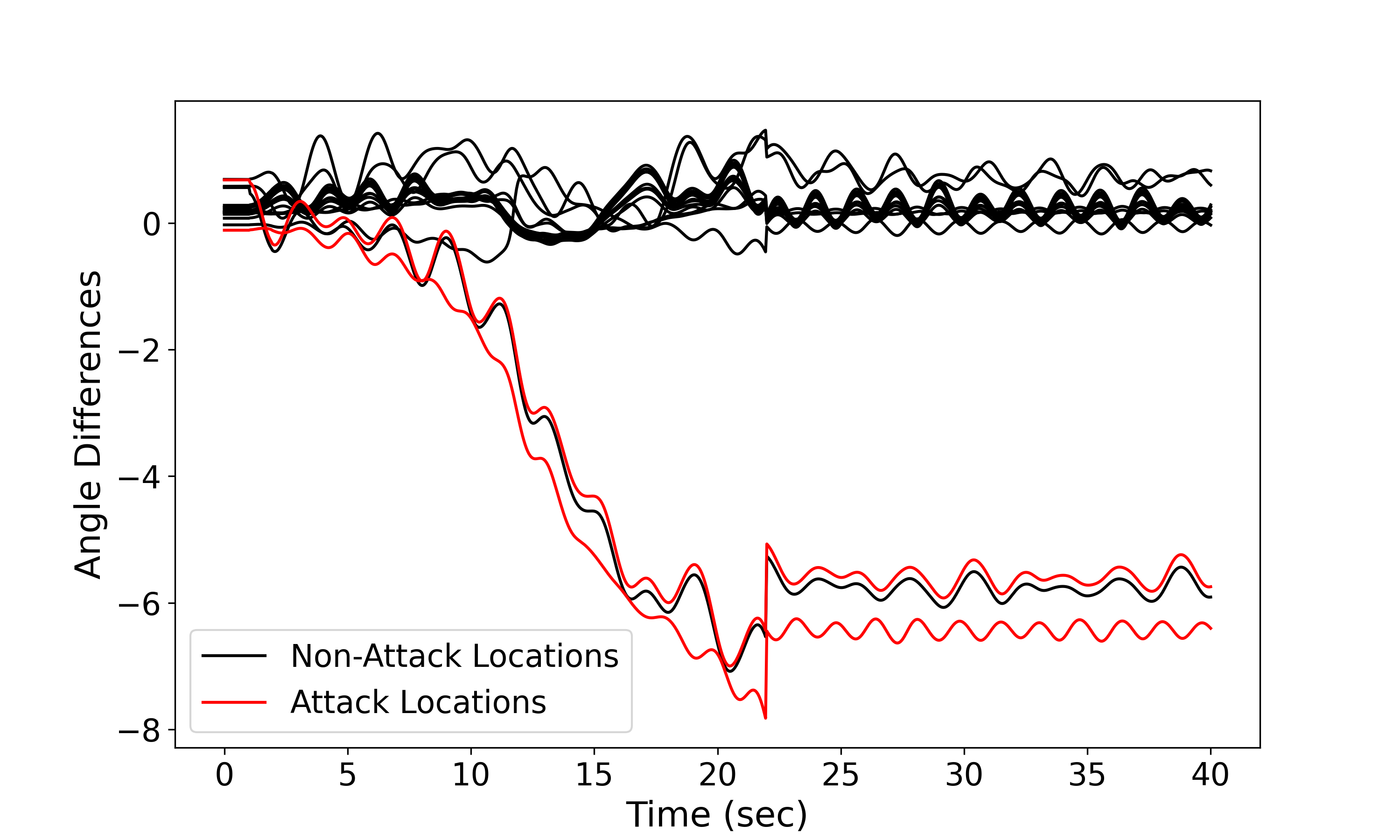}}
\caption{Step scenario with large attack magnitude with $c=1.03$. As expected, the large attack magnitude results in larger deviations in angle differences compared to the small attack magnitude scenario.}
\label{sce48_step}
\end{figure}

 In this paper, we define the step attack to be a multiplicative distortion, whereby true sensor measurements are scaled by a fixed factor for the duration of the attack. For a true voltage angle measurement at time $t$, $\phi_v(t)$, undergoing a step attack, the spurious voltage angle measurement, $\tilde{\phi}_v(t)$ is given by,
\begin{align}
    \tilde{\phi}_v(t) &= c \phi_v(t), \hspace{0.3in} \forall t \in [t_1, t_2],  
\end{align}
where $c$ is a constant, and $t_1$ and $t_2$ represent the start and end times of the cyber-attack. Two examples with voltage angle differences can be seen in Fig. \ref{sce46_step} and Fig. \ref{sce48_step}, with small and large attack magnitudes respectively. Please note that our angle data are normalized, therefore no unit for the angle difference. For all subsequent figures similar to these two, although the system consists of 68 buses, we will only be presenting data for the generator buses and the buses experiencing load changes. It is evident from Fig. \ref{sce48_step} that another bus exhibits patterns similar to those of the attacked buses, complicating the detection task.

\subsubsection{Poisoning}
\par We take the poison attack to be an additive attack whereby realisations of a random variable $C \sim N(\mu_{C}, \sigma_{C}^2)$, given by $c$, are added to individual angle measurements, such that,
\begin{align}
    \tilde{\phi}_v(t) &= \phi_v(t) + c, \hspace{0.3in} \forall t \in [t_1, t_2].
\end{align}

\begin{figure}[htbp]
\centerline{\includegraphics[width=0.95\columnwidth]{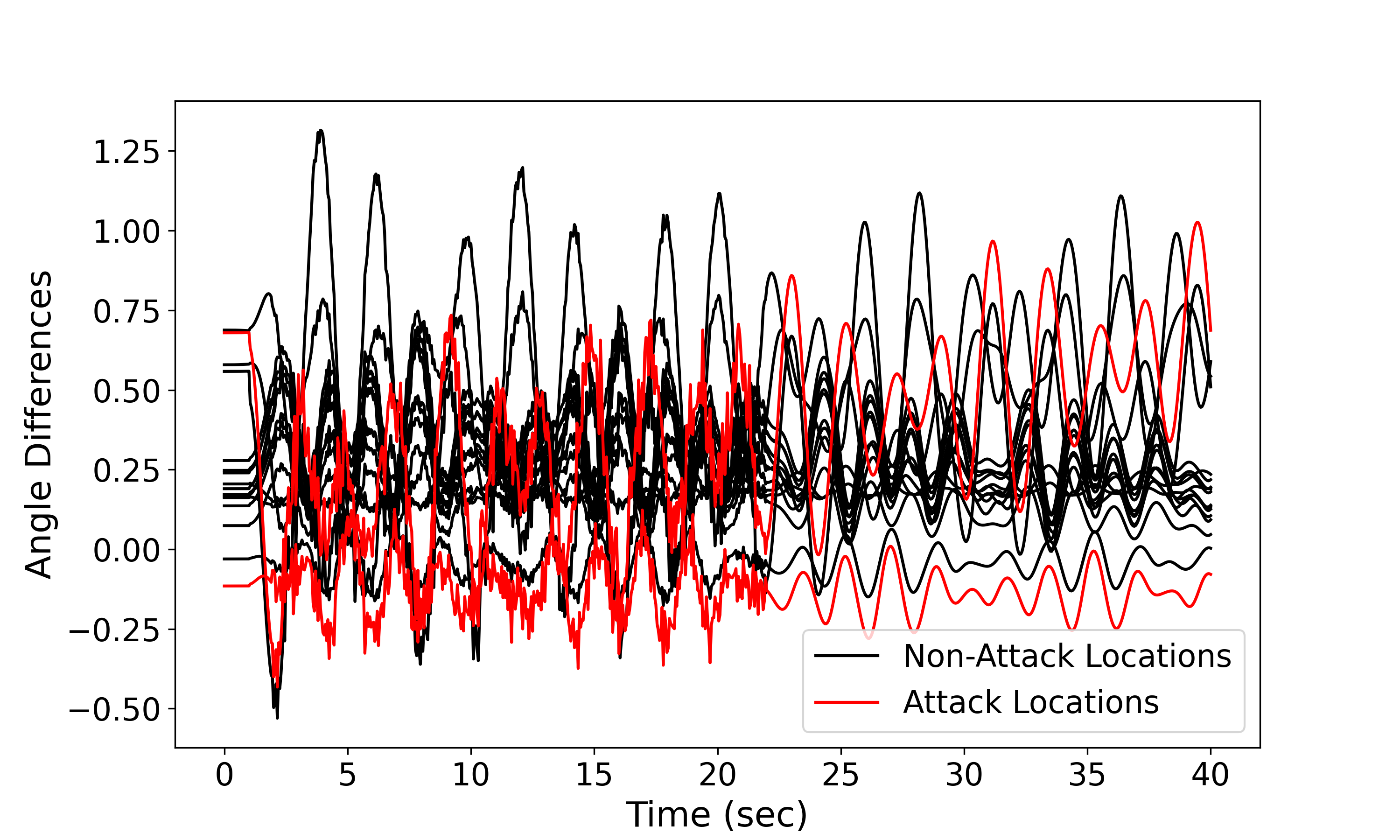}}
\caption{Poisoning scenario with large attack magnitude with $\mu_C = 0$ and $\sigma_C = 0.08$.}
\label{sce48_pois}
\end{figure}

An example with voltage angle differences can be seen in Fig. \ref{sce48_pois} with large attack magnitude.
%
%
\subsubsection{Ramp}
\par
\begin{figure}[htbp]
\centerline{\includegraphics[width=0.95\columnwidth]{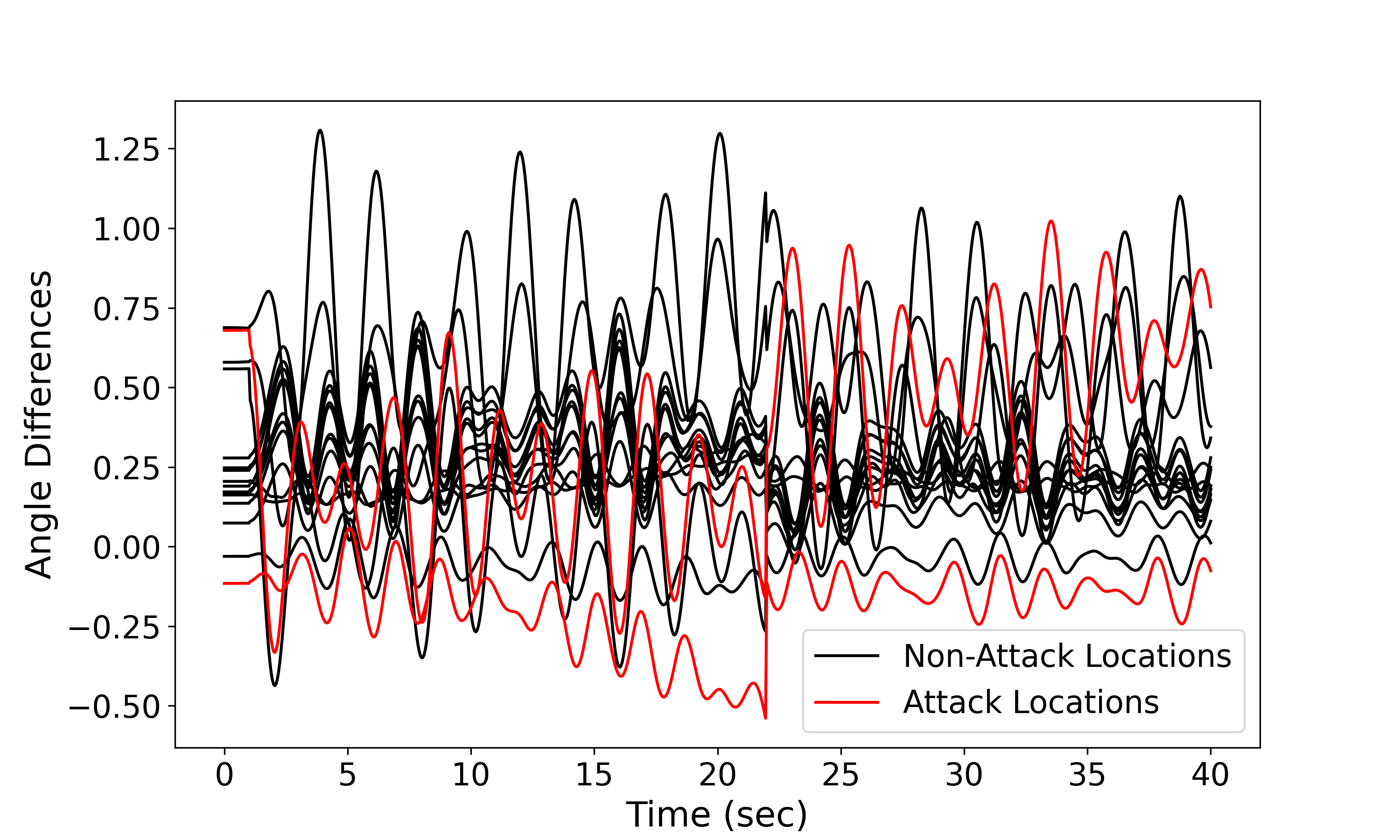}}
\caption{Ramp scenario with small attack magnitude, with $m=0.000007
$. The attack results in gradual divergence of angle differences during the attack.}
\label{sce43_ramp}
\end{figure}

\begin{figure}[htbp]
\centerline{\includegraphics[width=0.95\columnwidth]{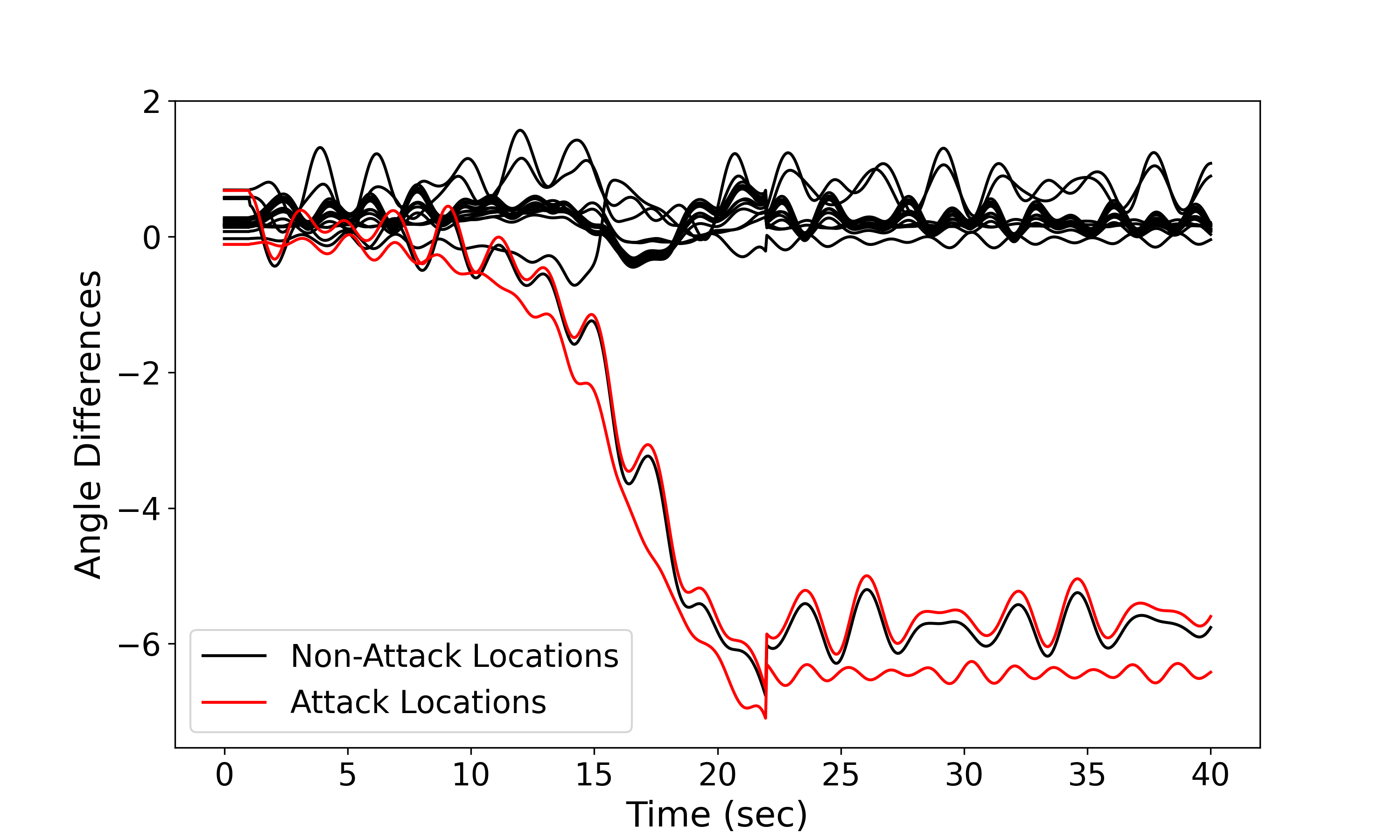}}
\caption{Ramp scenario with large attack magnitude, with $m=0.00007
$. As expected the large attack magnitude results in larger deviations in angle differences compared to the small attack magnitude scenario.}
\label{sce48_ramp}
\end{figure}

 In this work, we consider a ramp attack to be a scaling attack, whereby the function $c(t) = 1 + m\Delta t$ is multiplied to $\phi_v(t)$, where $m$ represents a preset gradient and $\Delta t = t - t_1$. The resulting spurious voltage angle measurements are given by,
\begin{align}
    \tilde{\phi}_v(t) &= c(t)\phi_v(t), \hspace{0.3in} \forall t \in [t_1, t_2],
    \label{eq:multiplicative_attack}
\end{align}
Two examples with voltage angle differences can be seen in Figs. \ref{sce43_ramp} and \ref{sce48_ramp}, with small and large attack magnitudes, respectively. As in Fig. \ref{sce48_step}, Fig. \ref{sce48_ramp} also presents a situation where another bus exhibits patterns similar to those of the attacked buses, making the detection task non-trivial.
\subsubsection{Riding the Wave}
\par
\begin{figure}[htbp]
\centerline{\includegraphics[width=0.95\columnwidth]{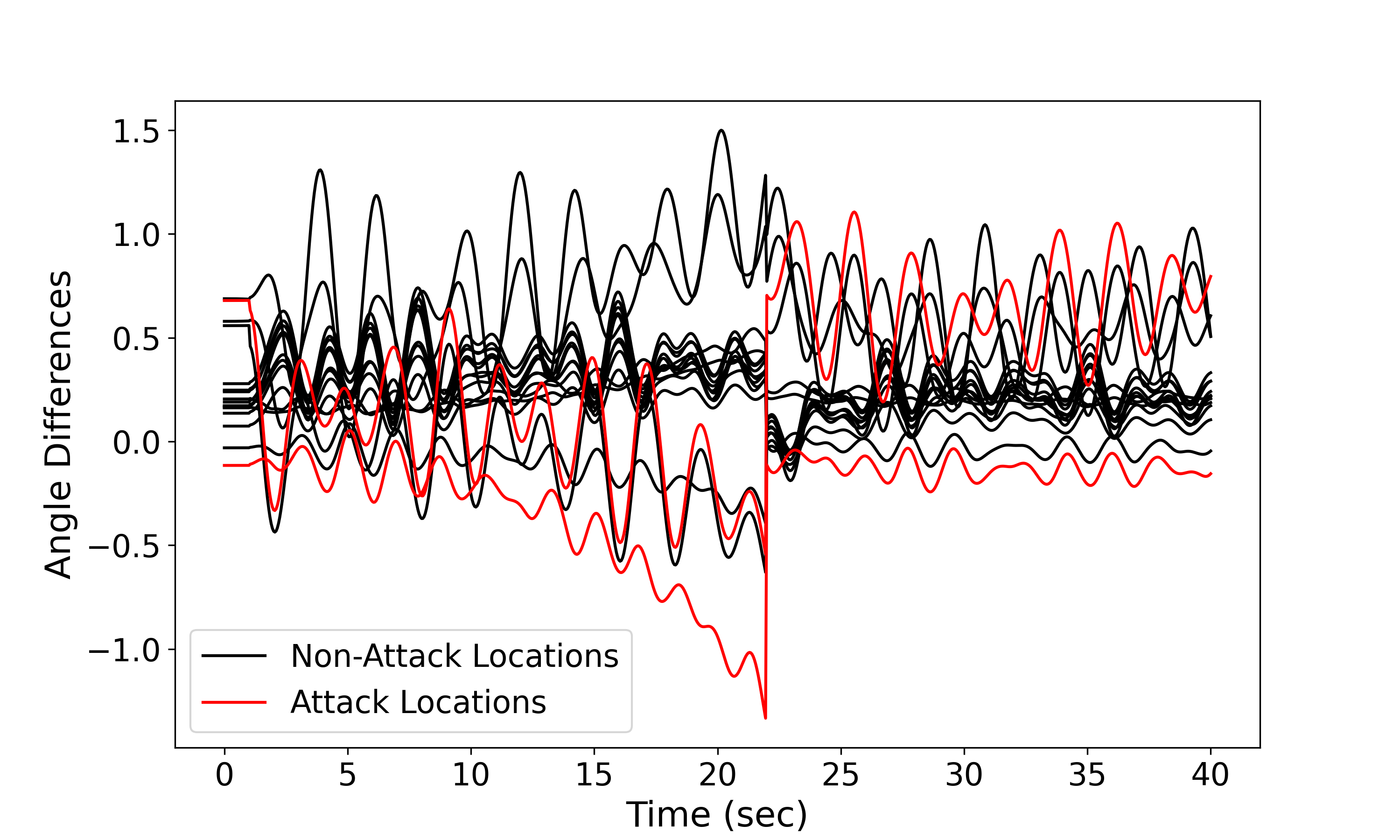}}
\caption{RTW scenario with small attack magnitude, with $\beta = 0.000325$.}
\label{sce43_RTW}
\end{figure}

\begin{figure}[htbp]
\centerline{\includegraphics[width=0.95\columnwidth]{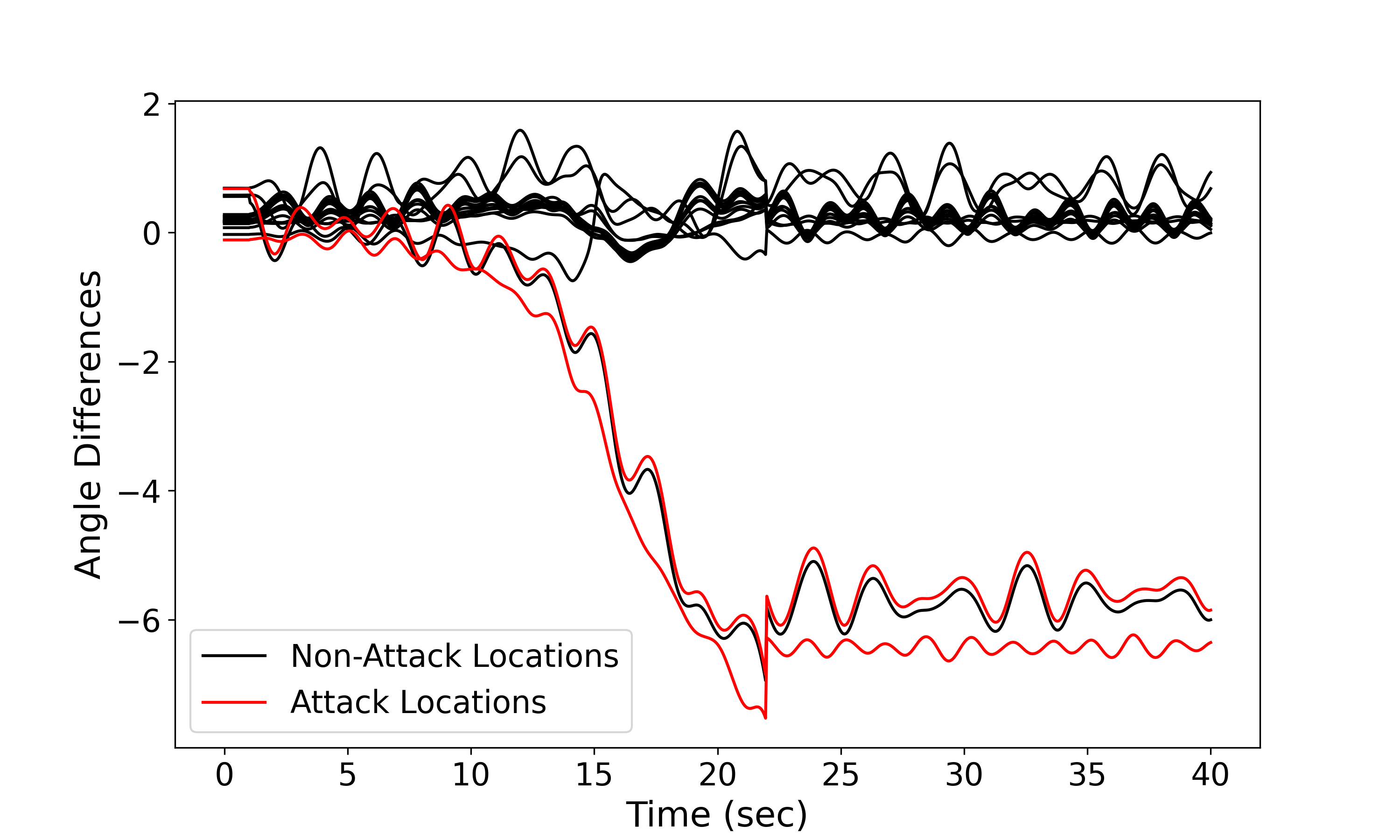}}
\caption{RTW scenario with large attack magnitude, with $\beta = 0.0015$.}
\label{sce48_RTW}
\end{figure}

 We define the RTW attack as a multiplicative attack that exploits the perturbation in the system resulting from a grid event, as in \cite{Nandanoori}. Specifically, this attack strategy defines a time-varying attack signal, $c(t)$, as,
\begin{align}
    c(t) &= \beta \Delta t (\phi_v(t) - \phi_v^{\text{nom}}), \hspace{0.15in} \forall t \in [t_1, t_2].
\end{align}
where $\beta$ is a constant and $\phi_v^{\text{nom}}$ is the nominal value of the voltage angle. Since this attack is multiplicative in nature, $\tilde{\phi}_v(t)$ is given by (\ref{eq:multiplicative_attack}). As the attack signal is proportional to the time elapsed since the start of the attack, this attack results in a delayed impact on the angle measurements. Furthermore, if the perturbation in the angle measurements due to a grid event is significant, this attack may also result in non-trivial changes in angle measurements, even beyond $t_2$. Consequently, control resources may be engaged to off-set these spurious disturbances. Two examples with voltage angle differences can be seen in Fig. \ref{sce43_RTW} and Fig. \ref{sce48_RTW}, with small and large attack magnitudes respectively. For Fig. \ref{sce48_RTW}, it is also a similar situation with Fig. \ref{sce48_step} that another bus exhibits patterns similar to those of the attacked buses.

In this study, we selected four distinct attack scenarios to comprehensively analyze the impact on power systems. These scenarios are categorized based on the proximity of the attacked buses to the bus where a load change occurs and the magnitude of the attack. Specifically, we have two scenarios where at least one of the attacked buses is near the load change bus, and two scenarios where all attacked buses are far away from the load change bus. For both proximity categories, we examine one scenario with a large attack magnitude and one with a small attack magnitude.

Within each of these scenarios, we simulate the four types of cyber-attacks: Step, Poisoning, Ramp, and RTW attacks. Among these, the RTW attack is expected to be the most challenging to detect due to its subtle nature. This particular attack type will be the primary focus of our results section.

\subsection{ML Methods}
\subsubsection{Conventional ML method}

In this approach, we leverage k-means clustering to identify under-attack behaviors within sliding windows of the time series data representing angle differences of the power system. We use 1 second of data as a window. The window data undergoes a transposition to rearrange its dimensions, positioning time series as individual samples for clustering. Utilizing the k-means algorithm configured to identify two distinct clusters, we engage in a clustering process on the transposed dataset. This step aims to differentiate between normal and potential anomalous operational patterns within the window. The core equation of the k-means clustering algorithm is the objective function that the algorithm seeks to minimize. This objective function is often referred to as the within-cluster sum of squares, which is defined as follows \cite{Lloyd}:
\begin{equation}
J = \sum_{k=1}^{K} \sum_{i \in C_k} \| \mathbf{x}_i - \mathbf{\mu}_k \|^2
\end{equation}
where:
\begin{itemize}
    \item $K$ is the number of clusters, set to be 2 in this study.
    \item $C_k$ is the set of points that belong to cluster $k$.
    \item $\mathbf{x}_i$ is a data point in cluster $k$.
    \item $\mathbf{\mu}_k$ is the centroid of cluster $k$.
    \item $\| \mathbf{x}_i - \mathbf{\mu}_k \|^2$ is the squared Euclidean distance between data point $\mathbf{x}_i$ and the centroid $\mathbf{\mu}_k$.
\end{itemize}

To evaluate the effectiveness of the clustering and the presence of distinct operational states, we compute the silhouette score for the clustered features. The silhouette score, ranging from -1 to 1, measures how similar an object is to its own cluster compared to other clusters. A high silhouette score suggests that the clusters are well separated and cohesive. The silhouette score \( s \) for a single sample is defined as follows \cite{Rousseeuw}:
\begin{equation}
s = \frac{b - a}{\max(a, b)}
\end{equation}
where:
\begin{itemize}
    \item $a$ is the mean intra-cluster distance (the average distance between the sample and all other points in the same cluster).
    \item $b$ is the mean nearest-cluster distance (the average distance between the sample and all points in the nearest cluster that the sample is not a part of).
\end{itemize}

Should the average silhouette score exceed a pre-established threshold (e.g., 0.8), indicating clear separation between clusters, we proceed to identify attacks. This involves determining the less populous of the two clusters, under the assumption that anomalous behaviors are less common. The indices of time series within this cluster are flagged as anomalies, suggesting deviations from typical operational patterns. Conversely, if the silhouette score does not meet the threshold, it indicates insufficient separation between clusters to reliably identify attacks.


\subsubsection{Deep learning-based method}
We also conduct anomaly detection through the integration of an autoencoder model with k-means clustering. We use 1 second of data as a window. For each window of the time series data, representing specific segments of operational metrics, we first construct an autoencoder. This autoencoder is designed with an input layer corresponding to the number of time series, followed by successive dense layers for encoding and decoding processes. The aim is to capture and reconstruct the operational data patterns. An early stopping mechanism is employed to prevent overfitting, using validation loss as a monitoring metric. The reconstruction error for an autoencoder is defined as the mean squared error between the input and the reconstructed output \cite{Hinton}:
\begin{equation}
\mathcal{L}(x, \hat{x}) = \| x - \hat{x} \|^2
\end{equation}
where:
\begin{itemize}
    \item $\mathcal{L}(x, \hat{x})$ is the reconstruction loss.
    \item $x$ is the original input.
    \item $\hat{x}$ is the reconstructed output.
    \item $\| x - \hat{x} \|^2$ represents the squared Euclidean distance between $x$ and $\hat{x}$.
\end{itemize}

The model training commences with the initial window's data, progressively incorporating additional data up to the current window to refine the autoencoder's ability to generalize across varying operational states. After training, we utilize the autoencoder to generate reconstructions of the current window's data. The deviation between the actual data and its reconstruction, quantified as the reconstruction error, serves as an indicator of anomalous patterns. We then average these errors across samples to determine a mean error for each time series, forming a basis for clustering.

Applying k-means clustering on these mean error values, with the aim to distinguish between normal and anomalous operational patterns, again results in two clusters. The differentiation is further validated by calculating the silhouette score on the reconstruction error like the approach described above. This combination of autoencoder-based feature extraction with k-means clustering provides a nuanced approach to identifying cyber-attacks.



\subsubsection{Graph neural network approaches}
In the realm of multivariate time series anomaly detection, traditional methods often struggle to capture the complex interdependencies and stochastic nature inherent in industrial datasets. Recent advancements, however, have introduced GNNs as potent tools for addressing these challenges. Specifically, the GAT and GDN represent innovative approaches that leverage the structural relationships between time series data points to enhance anomaly detection. GAT and GDN take ``normal data'' as the training data. When training GAT and GDN, we use five scenarios of simulated data without attack as the training data.

Both GAT and GDN utilizes graph attention layers where the attention coefficients \(\alpha_{ij}\) between nodes \(i\) and \(j\) in a graph attention network are calculated as follows \cite{Veličković}:

\begin{equation}
\alpha_{ij} = \frac{\exp\left(\text{LeakyReLU}\left(\mathbf{a}^T [\mathbf{W}\mathbf{h}_i \| \mathbf{W}\mathbf{h}_j]\right)\right)}{\sum_{k \in \mathcal{N}_i} \exp\left(\text{LeakyReLU}\left(\mathbf{a}^T [\mathbf{W}\mathbf{h}_i \| \mathbf{W}\mathbf{h}_k]\right)\right)}
\end{equation}

where:
\begin{itemize}
    \item \(\mathbf{h}_i\) and \(\mathbf{h}_j\) are the input features of nodes \(i\) and \(j\), respectively.
    \item \(\mathbf{W}\) is the weight matrix applied to the input features.
    \item \(\mathbf{a}\) is the attention mechanism's weight vector.
    \item \(\|\) denotes concatenation.
    \item \(\mathcal{N}_i\) represents the set of neighbors of node \(i\).
    \item \(\text{LeakyReLU}\) is the Leaky Rectified Linear Unit activation function.
\end{itemize}

The final output features for node \(i\) are computed as a weighted sum of the transformed neighboring features:

\begin{equation}
\mathbf{h}'_i = \sigma \left( \sum_{j \in \mathcal{N}_i} \alpha_{ij} \mathbf{W}\mathbf{h}_j \right)
\end{equation}

where:
\begin{itemize}
    \item \(\mathbf{h}'_i\) is the output feature for node \(i\).
    \item \(\sigma\) is a non-linear activation function, such as ReLU.
\end{itemize}

The GAT focuses on learning the intricate dependencies dependencies and interactions among different time series across both temporal space and feature dimensions, which are often neglected by traditional anomaly detection techniques. In this approach, each time series is treated as a node in a graph, and the relationships between the time series are represented as edges. It employs an attention mechanism that dynamically assigns different weights to the edges based on their importance. This attention mechanism enables the model to focus on the most relevant time series when making predictions. Anomaly detection in GAT is based on the reconstruction and forecasting errors. The model is trained to reconstruct the normal patterns of the time series data and predict future values. During the detection phase, anomalies are identified by measuring the discrepancies between the actual values and the reconstructed or predicted values. High reconstruction or forecasting errors indicate potential anomalies, as they suggest that the observed data deviates significantly from the learned normal patterns. 

The GDN proposes a structure learning strategy with also the attention mechanisms, facilitating a deeper understanding of inter-sensor relationships. The core of GDN lies in its graph structure learning and graph-attention based forecasting. The attention mechanism operates by considering the local context of each node, aggregating information from its neighboring nodes. This allows the GDN to highlight which time series and interactions are most critical for identifying anomalies at any given time step. Anomaly detection in GDN is primarily based on deviation scores. The model extracts attention-based features from the time series data using the learned graph structure, then computes a deviation score for each node, which measures how much the current observation deviates from the expected normal pattern learned during training. These deviation scores are then aggregated using the maximum function to produce an overall anomaly score \cite{Deng}:

\begin{equation}
A(t) = \max_{i} a_i(t)
\end{equation}

When training GAT and GDN, the settings are similar as we utilize a 1 second window for both models. Generally, GDN training is computationally efficient, typically completing within 10 minutes using a single NVIDIA Tesla P100 GPU. In contrast, GAT training may extend beyond 10 minutes but generally completes within 30 minutes.

These methodologies not only claim superior detection capabilities compared to earlier models but also improve interpretability. The interpretability is primarily derived from the analysis of forecasting or reconstruction errors. By examining these errors for each bus within the power grid, we can potentially identify which specific buses are experiencing anomalous behavior indicative of an attack. In the results section, we delve into these findings.

\section{Results}

\subsection{Detection}
\begin{figure}[htbp]
\centerline{\includegraphics[width=0.95\columnwidth]{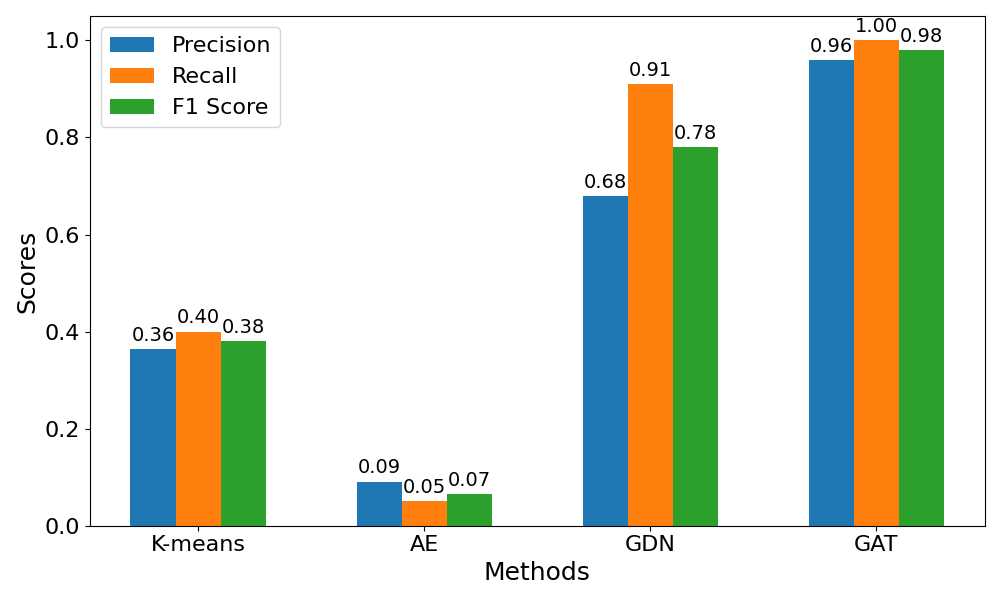}}
\caption{RTW scenario small attack (load change and attack near). GAT and GDN outperformed k-means and autoencoder, with GAT achieving nearly perfect detection results.}
\label{sce43_rtw_detection}
\end{figure}

\begin{figure}[htbp]
\centerline{\includegraphics[width=0.95\columnwidth]{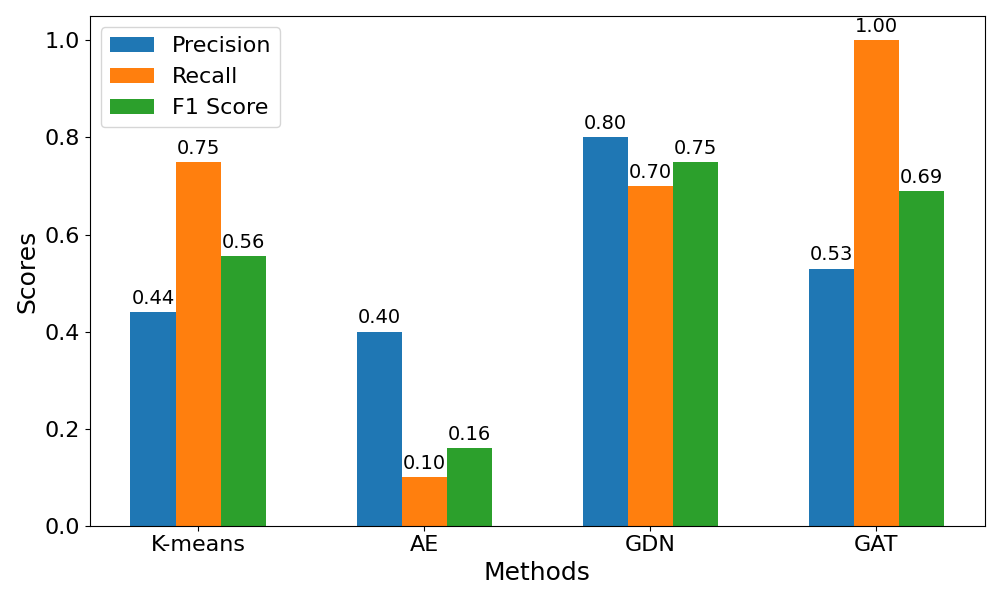}}
\caption{RTW scenario large attack (load change and attack near). GAT and GDN outperformed k-means and autoencoder, with GDN surpassing GAT on this occasion, although both encountered some difficulties.}
\label{sce48_rtw_detection}
\end{figure}

Because RTW attack is the most difficult one to detect, we first focus on the performance of all methods on RTW attack. We present the performance metrics, specifically precision, recall, and F1 scores. Precision focuses on how many of the predicted attacks are actual attacks, making it important if false alarms are costly. Recall tells you how many of the actual attacks were correctly identified, which is critical when missing attacks is costly. F1 score provides a balance between precision and recall, especially when you want to ensure both false positives and false negatives are minimized. We compare the detection results obtained from GAT and GDN with those from k-means clustering and autoencoder, as illustrated in Figs. \ref{sce43_rtw_detection} and \ref{sce48_rtw_detection}. Our findings indicate that both GAT and GDN outperform the conventional machine learning methods when the attack magnitude is small. However, when the attack magnitude is large and certain angle differences shift to a different state, GAT encounters difficulties. Despite this, GAT and GDN still maintain better performance compared to k-means and autoencoder. For the following results, we focus on comparing GAT and GDN since they are most of the time superior than k-means clustering and autoencoder.

The detailed detection results using the GAT and GDN models are provided in Tables \ref{tab1} to \ref{tab4}. Our analysis reveals that, in most cases, the GAT model exhibits superior performance over the GDN model, often detecting attacks with near-perfect accuracy. However, there are instances where the GAT model struggles, and the GDN model demonstrates better performances. These challenging cases are usually similar scenarios as the one illustrated in Figs. \ref{sce48_step}, \ref{sce48_ramp} and \ref{sce48_RTW}, where the angle differences of certain buses transition to a different state and remain there. In these scenarios, the GDN model is able to determine the cessation of the attack, whereas the GAT model continues to indicate the presence of an attack.

\begin{table}[htbp]
\caption{Scenario small attack (load change and attack near)}
\centering
\begin{tabular}{|c|c|c|c|c|}
\hline
 & Poison & Ramp & RTW & Step \\ \hline
GDN & 
\begin{tabular}[c]{@{}r@{}}F1: 0.72\\ prec: 0.58\\ recall: 0.94\end{tabular} & 
\begin{tabular}[c]{@{}r@{}}F1: 0.78\\ prec: 0.66\\ recall: 0.93\end{tabular} & 
\begin{tabular}[c]{@{}r@{}}F1: 0.78\\ prec: 0.68\\ recall: 0.91\end{tabular} & 
\begin{tabular}[c]{@{}r@{}}F1: 0.81\\ prec: 0.71\\ recall: 0.94\end{tabular} \\ \hline
GAT & 
\begin{tabular}[c]{@{}r@{}}F1: 1.00\\ prec: 1.00\\ recall: 1.00\end{tabular} & 
\begin{tabular}[c]{@{}r@{}}F1: 1.00\\ prec: 1.00\\ recall: 1.00\end{tabular} & 
\begin{tabular}[c]{@{}r@{}}F1: 0.98\\ prec: 0.96\\ recall: 1.00\end{tabular} & 
\begin{tabular}[c]{@{}r@{}}F1: 0.97\\ prec: 0.95\\ recall: 1.00\end{tabular} \\ \hline
\end{tabular}
\label{tab1}
\end{table}

\begin{table}[htbp]
\caption{Scenario large attack (load change and attack near)}
\centering
\begin{tabular}{|c|c|c|c|c|}
\hline
 & Poison & Ramp & RTW & Step \\ \hline
GDN & 
\begin{tabular}[c]{@{}r@{}}F1: 0.73\\ prec: 0.60\\ recall: 0.91\end{tabular} & 
\begin{tabular}[c]{@{}r@{}}F1: 0.69\\ prec: 0.75\\ recall: 0.64\end{tabular} & 
\begin{tabular}[c]{@{}r@{}}F1: 0.75\\ prec: 0.80\\ recall: 0.70\end{tabular} & 
\begin{tabular}[c]{@{}r@{}}F1: 0.77\\ prec: 0.85\\ recall: 0.69\end{tabular} \\ \hline
GAT & 
\begin{tabular}[c]{@{}r@{}}F1: 1.00\\ prec: 1.00\\ recall: 1.00\end{tabular} & 
\begin{tabular}[c]{@{}r@{}}F1: 0.69\\ prec: 0.52\\ recall: 1.00\end{tabular} & 
\begin{tabular}[c]{@{}r@{}}F1: 0.69\\ prec: 0.53\\ recall: 1.00\end{tabular} & 
\begin{tabular}[c]{@{}r@{}}F1: 0.69\\ prec: 0.53\\ recall: 1.00\end{tabular} \\ \hline
\end{tabular}
\label{tab2}
\end{table}

\begin{table}[htbp]
\caption{Scenario small attack (load change and attack far)}
\centering
\begin{tabular}{|c|c|c|c|c|}
\hline
 & Poison & Ramp & RTW & Step \\ \hline
GDN & 
\begin{tabular}[c]{@{}r@{}}F1: 0.72 \\ prec: 0.58 \\ recall: 0.95\end{tabular} & 
\begin{tabular}[c]{@{}r@{}}F1: 0.75 \\ prec: 0.65 \\ recall: 0.88\end{tabular} & 
\begin{tabular}[c]{@{}r@{}}F1: 0.83 \\ prec: 0.82 \\ recall: 0.84\end{tabular} & 
\begin{tabular}[c]{@{}r@{}}F1: 0.82 \\ prec: 0.80 \\ recall: 0.84\end{tabular} \\ \hline
GAT & 
\begin{tabular}[c]{@{}r@{}}F1: 1.00 \\ prec: 1.00 \\ recall: 1.00\end{tabular} & 
\begin{tabular}[c]{@{}r@{}}F1: 1.00 \\ prec: 1.00 \\ recall: 1.00\end{tabular} & 
\begin{tabular}[c]{@{}r@{}}F1: 1.00 \\ prec: 1.00 \\ recall: 1.00\end{tabular} & 
\begin{tabular}[c]{@{}r@{}}F1 : 1.00 \\ prec: 1.00 \\ recall: 1.00\end{tabular} \\ \hline
\end{tabular}
\label{tab3}
\end{table}

\begin{table}[htbp]
\caption{Scenario large attack (load change and attack far)}
\centering
\begin{tabular}{|c|c|c|c|c|}
\hline
 & Poison & Ramp & RTW & Step \\ \hline
GDN & 
\begin{tabular}[c]{@{}r@{}}F1: 0.72 \\ prec: 0.68 \\ recall: 0.77\end{tabular} & 
\begin{tabular}[c]{@{}r@{}}F1: 0.80 \\ prec: 0.82 \\ recall: 0.78\end{tabular} & 
\begin{tabular}[c]{@{}r@{}}F1: 0.91 \\ prec: 0.92 \\ recall: 0.90\end{tabular} & 
\begin{tabular}[c]{@{}r@{}}F1: 0.90 \\ prec: 0.84 \\ recall: 0.96\end{tabular} \\ \hline
GAT & 
\begin{tabular}[c]{@{}r@{}}F1: 1.00 \\ prec: 1.00 \\ recall: 1.00\end{tabular} & 
\begin{tabular}[c]{@{}r@{}}F1: 0.69 \\ prec: 0.52 \\ recall: 1.00\end{tabular} & 
\begin{tabular}[c]{@{}r@{}}F1: 0.69 \\ prec: 0.53 \\ recall: 1.00\end{tabular} & 
\begin{tabular}[c]{@{}r@{}}F1: 0.69 \\ prec: 0.52 \\ recall: 1.00\end{tabular} \\ \hline
\end{tabular}
\label{tab4}
\end{table}

\subsection{Location}
While detecting attacks on buses is important, it is more crucial for operators to know the precise locations of these buses. In GDN, we utilize the resulting deviation scores from detection to investigate this while in GAT, we combine forecast and reconstruction losses to find the attacked buses.

\begin{figure}[htbp]
\centerline{\includegraphics[width=0.95\columnwidth]{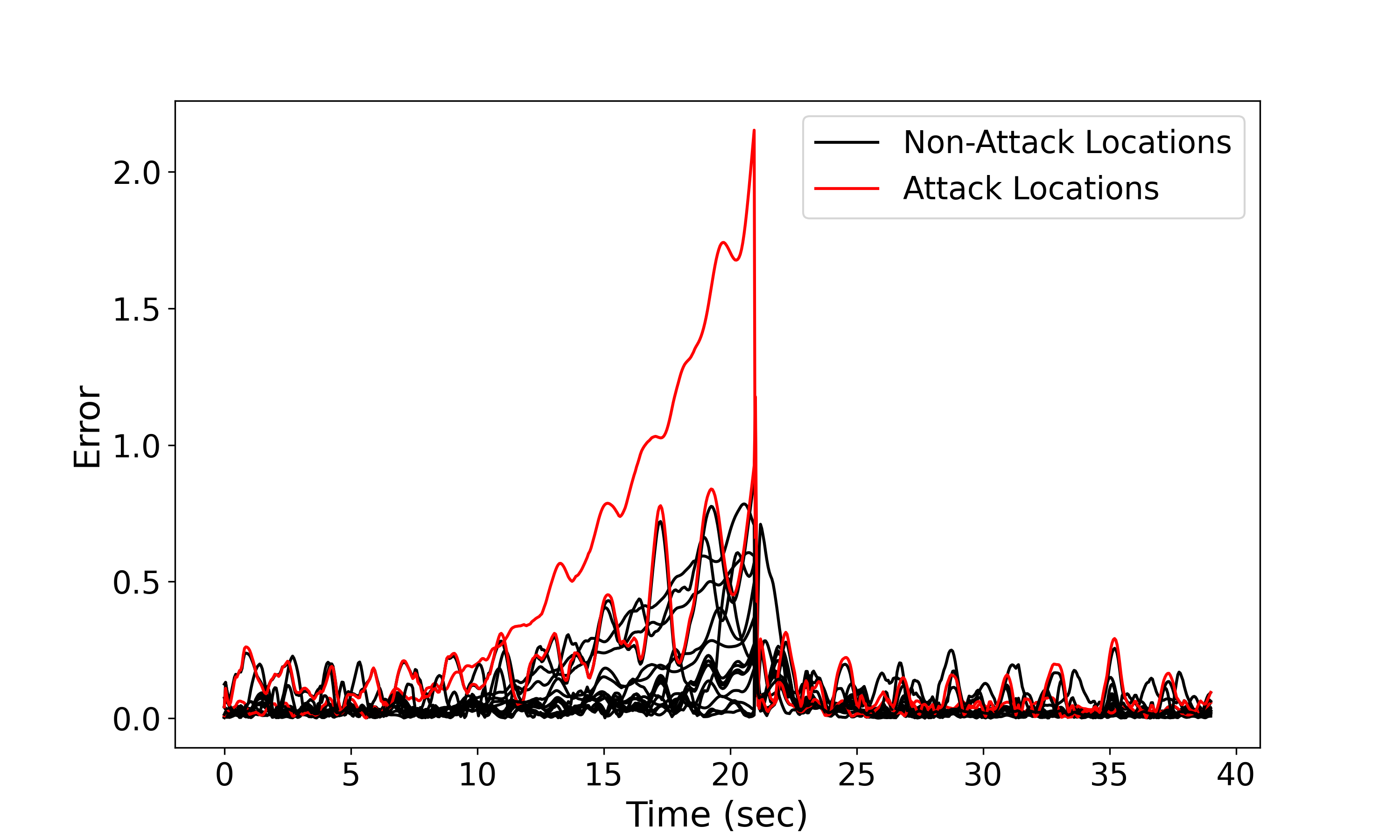}}
\caption{RTW scenario small attack (load change near), GAT losses. One of the attacked buses is obscured by losses with other buses, while the other attacked bus stands out only for a period of time during the attack period.}
\label{sce43_rtw_GAT_loc}
\end{figure}

\begin{figure}[htbp]
\centerline{\includegraphics[width=0.95\columnwidth]{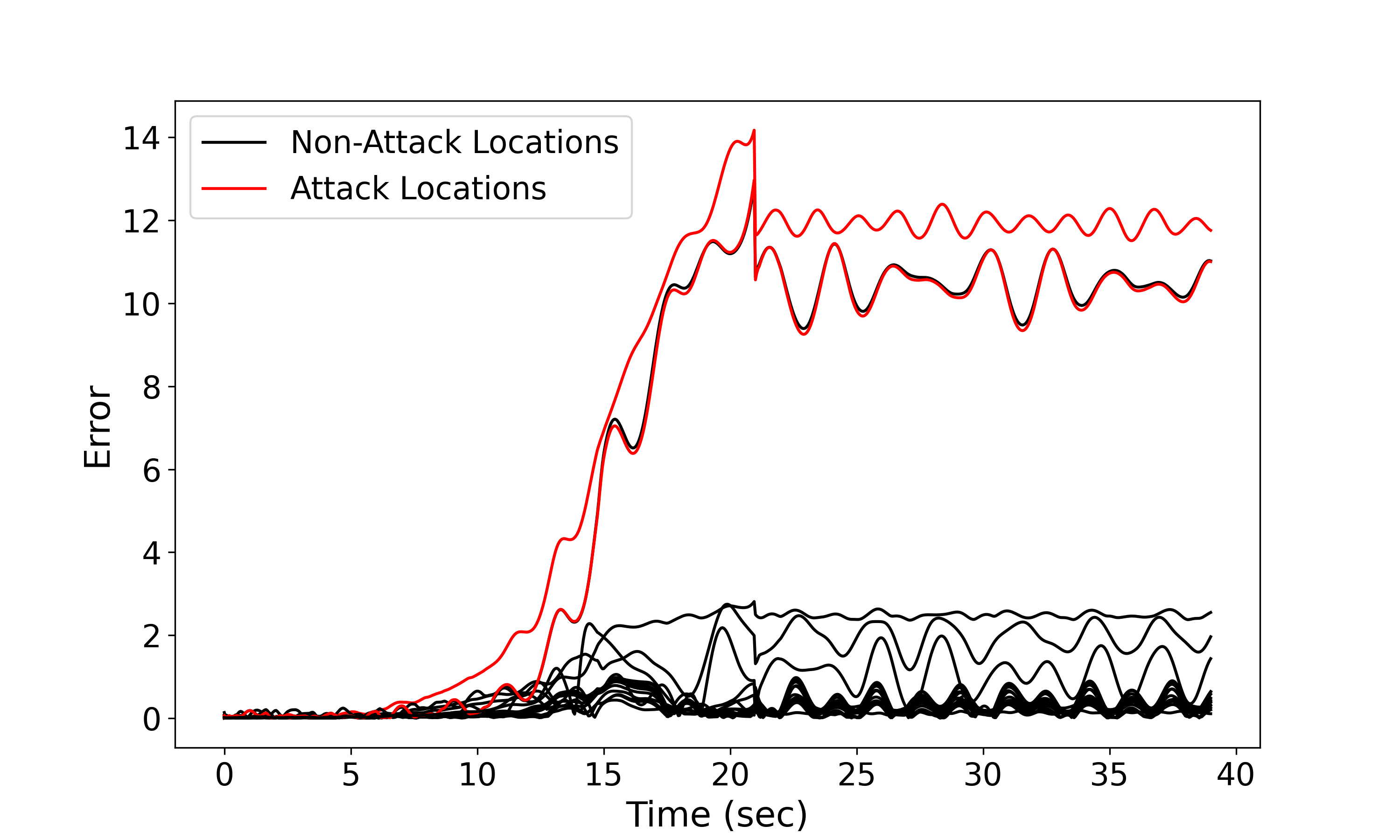}}
\caption{RTW scenario large attack (load change near), GAT losses. The two attacked buses do stand out; however, another non-attacked bus exhibits similar behavior, complicating the location identification task.}
\label{sce48_rtw_GAT_loc}
\end{figure}

\begin{figure}[htbp]
\centerline{\includegraphics[width=0.95\columnwidth]{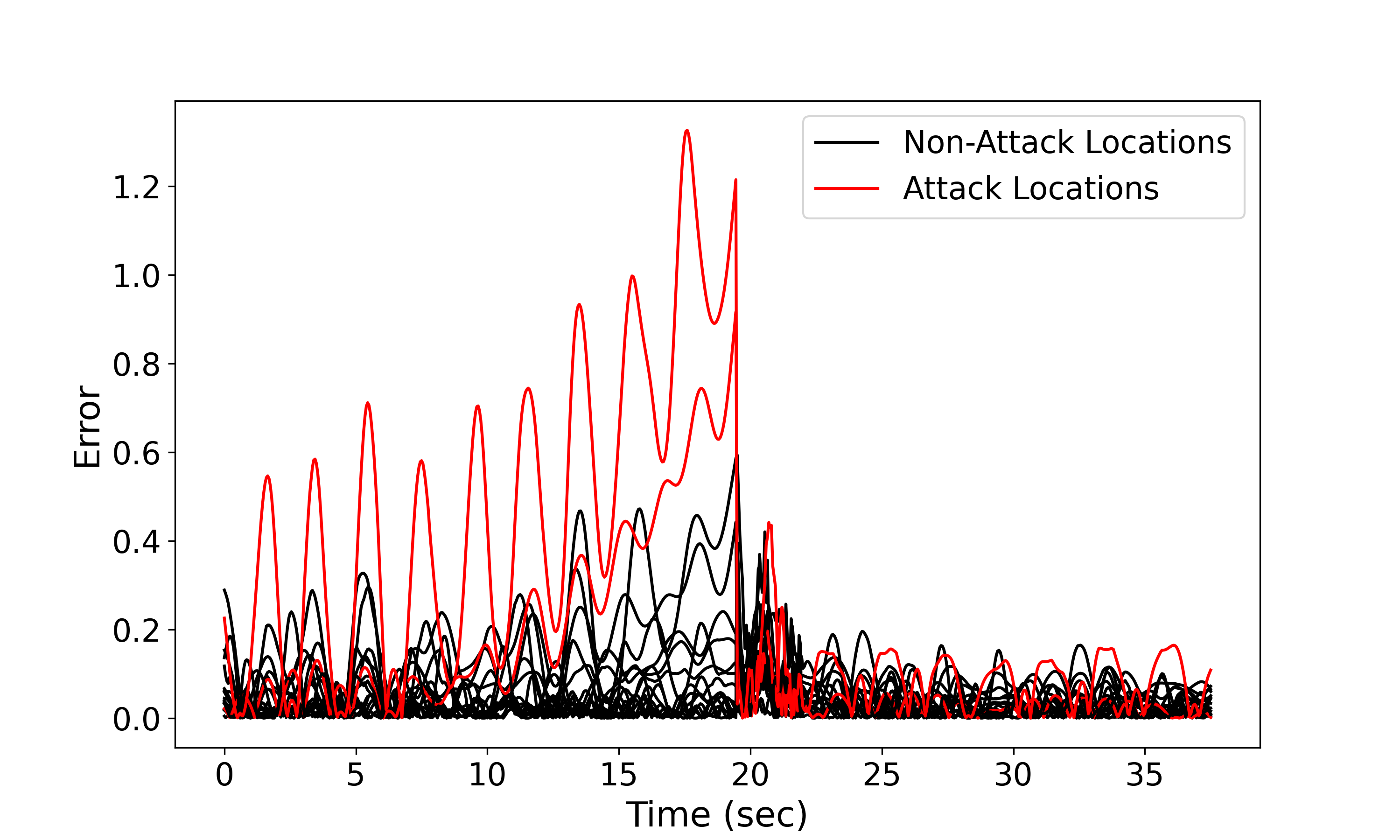}}
\caption{RTW scenario small attack (load change near), GDN losses. One of the attacked buses is obscured by losses with other buses, while the other attacked bus stands out only intermittently during the attack period.}
\label{sce43_rtw_GDN_loc}
\end{figure}

\begin{figure}[htbp]
\centerline{\includegraphics[width=0.95\columnwidth]{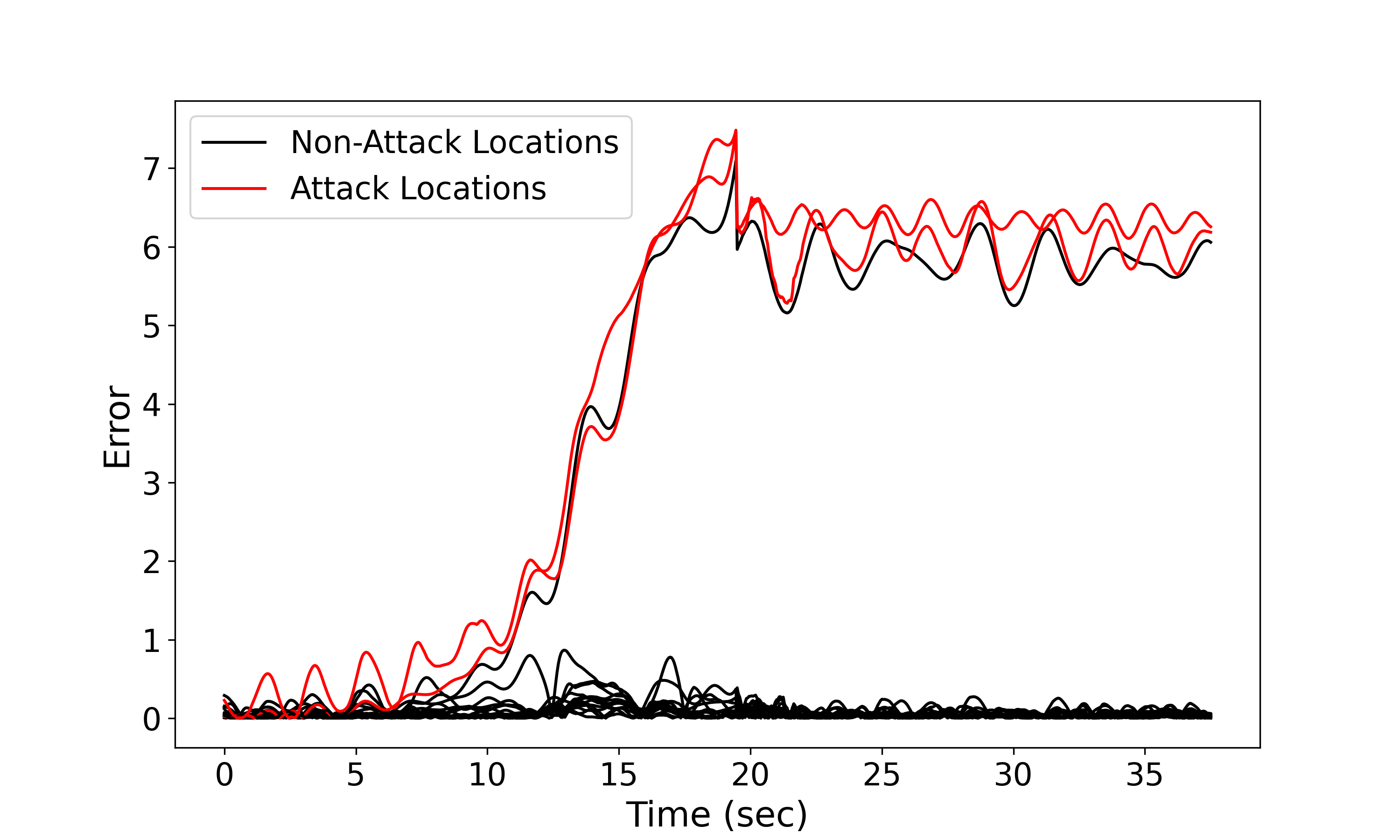}}
\caption{RTW scenario large attack, GDN losses. The two attacked buses do stand out; however, another non-attacked bus exhibits similar behavior, complicating the location identification task.}
\label{sce48_rtw_GDN_loc}
\end{figure}

\begin{figure}[htbp]
\centerline{\includegraphics[width=0.95\columnwidth]{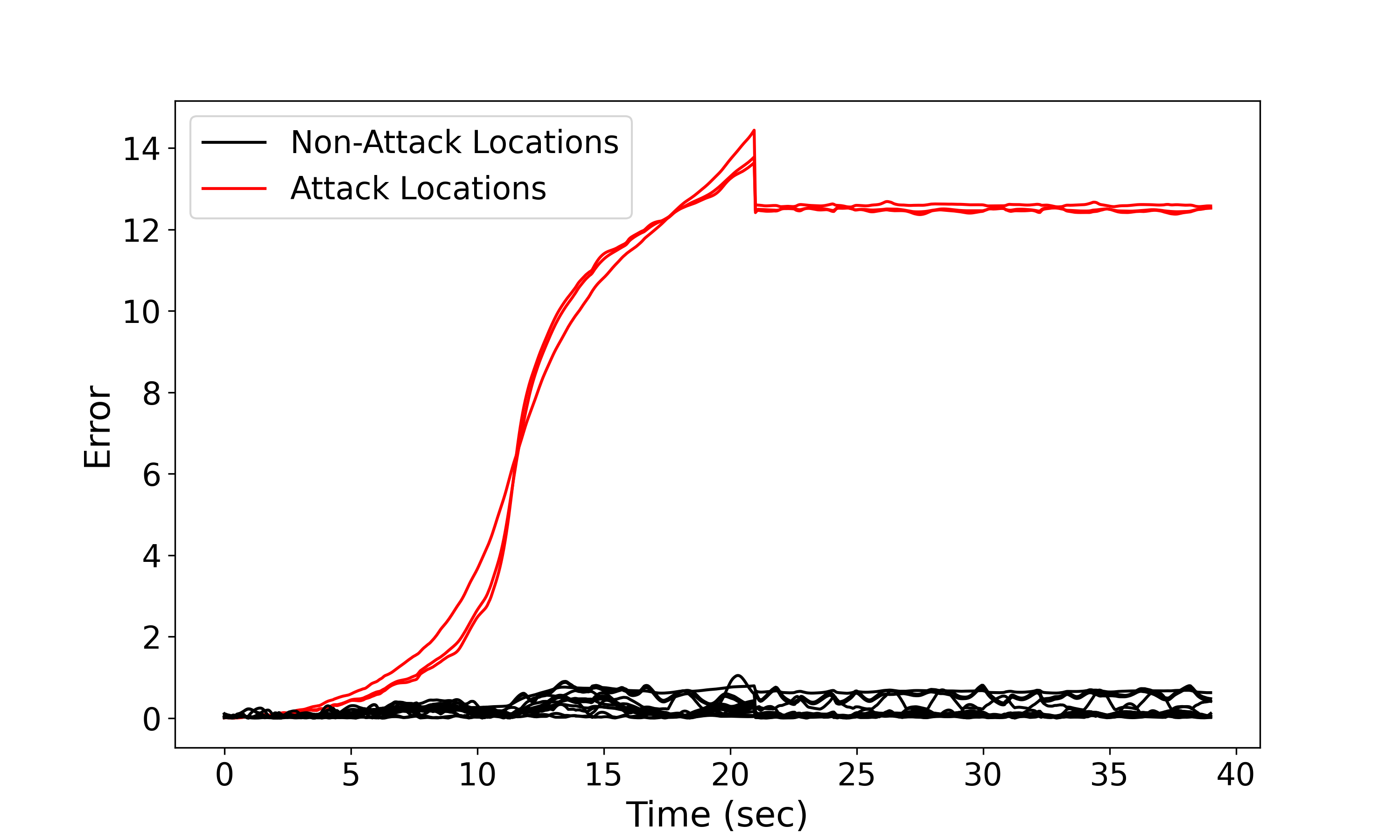}}
\caption{Step scenario large attack (load change far) GAT losses. The three attacked buses are distinctly different from the rest of the buses.}
\label{sce55_step_GAT_loc}
\end{figure}

\begin{figure}[htbp]
\centerline{\includegraphics[width=0.95\columnwidth]{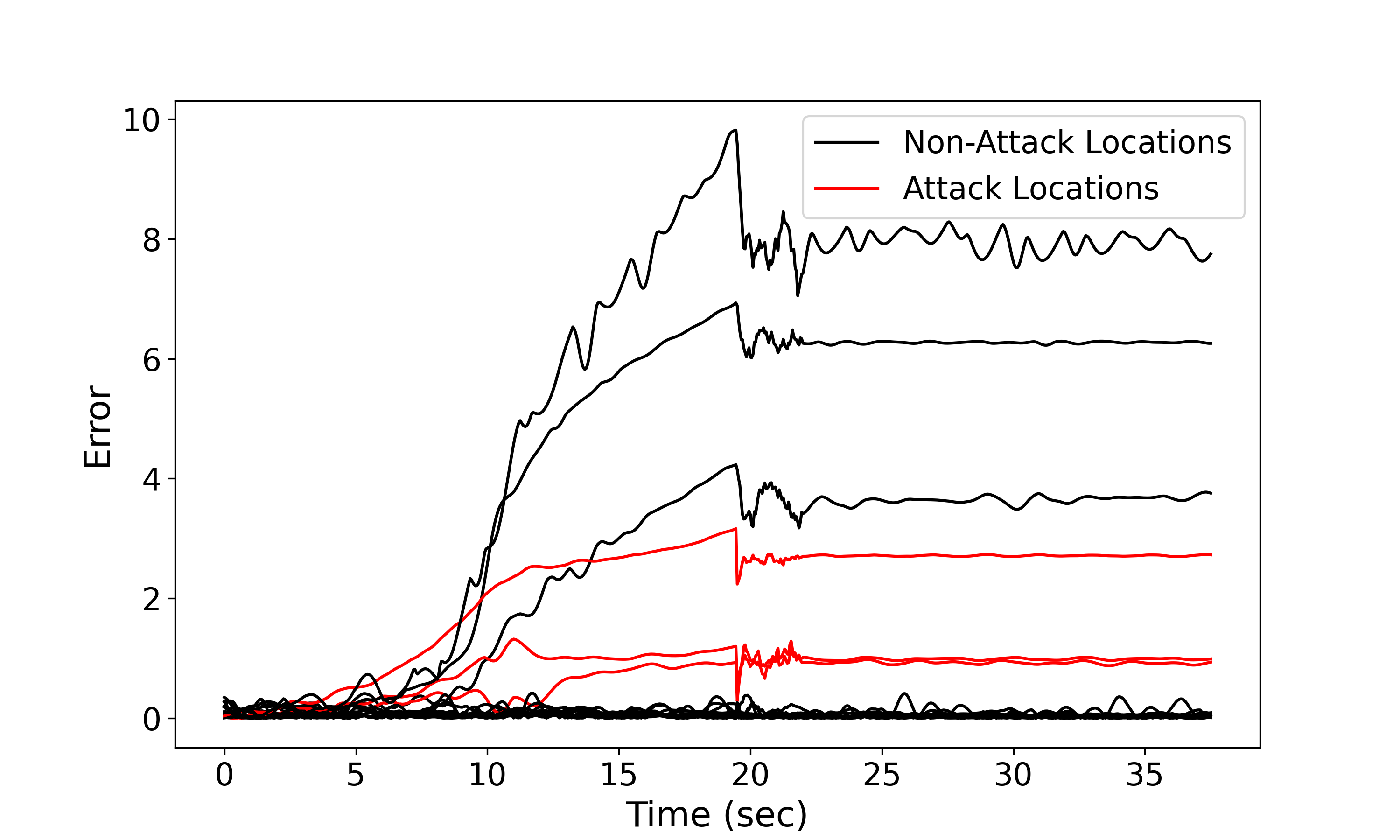}}
\caption{Step scenario large attack (load change far) GDN losses. The three attacked buses are not distinguishable from the other buses, as some non-attacked buses exhibit higher losses.}
\label{sce55_step_GDN_loc}
\end{figure}

Figs. \ref{sce43_rtw_GAT_loc} and \ref{sce48_rtw_GAT_loc} illustrate the combined forecast and reconstruction losses for all buses from GAT, where the buses under attack are represented by red lines. Figs. \ref{sce43_rtw_GDN_loc} and \ref{sce48_rtw_GDN_loc} illustrate the deviation scores from GDN. Analyzing these figures reveals that distinguishing the buses under attack remains a challenging task for RTW attack. The overlapping loss patterns and subtle deviations make it difficult to accurately identify the affected buses solely based on these metrics. In contrast, Fig. \ref{sce55_step_GAT_loc} presents the results for a simpler attack scenario, the Step attack. In this case, it is evident that the attacked buses are clearly distinguishable from the others based on the combined forecasting and reconstruction losses from GAT. However, the deviation scores obtained from the GDN do not effectively distinguish buses under attack from normal buses for the Step attack, as indicated in Fig. \ref{sce55_step_GDN_loc}.


\section{Discussion and Conclusion}

In this paper, we analyze the ability of machine learning approaches for detecting and localizing cyber-attacks on power systems. Our results demonstrate that GNN-based methods, specifically GAT and GDN, outperform conventional machine learning methods such as k-means clustering and autoencoder in detection. GNN-based methods also show promise in localizing attacks for simple scenarios. However, these methods encounter difficulties in more complex scenarios such as RTW attack.

In our detection analysis, GAT generally outperforms GDN, except in instances where the angle differences of some buses shift to a new state. In these cases, GAT struggles to detect the end of the attack, suggesting that GAT might be more sensitive to anomalies than GDN. For localization, both GAT and GDN face challenges with RTW attacks, the most difficult type of attack. Conversely, in simpler scenarios, such as the Step attack, GAT effectively distinguishes between buses under attack and those not affected.

Overall, GNN-based methods exhibit potential for cyber-attack detection and localization in power systems. Nevertheless, further refinement is necessary to enhance their performance for challenging scenarios. Future work should focus on tailoring these methods to better address the specific needs of power system security.

\bibliographystyle{IEEEtran}
\bibliography{references}



\vspace{12pt}

\end{document}